\documentclass[preprint,showpacs,preprintnumbers,amsmath,amssymb]{revtex4}

\usepackage{graphicx}
\usepackage{dcolumn}
\usepackage{bm}
\begin{document}


\title{Point-Contact Spectroscopy of the Heavy-Fermion Superconductor
CeCoIn$_5$}

\author{W. K. Park}
\author{L. H. Greene}
\affiliation{Department of Physics and Frederick Seitz Materials Research
Laboratory, University of Illinois at Urbana-Champaign, Urbana, Illinois 61801,
USA}

\author{J. L. Sarrao}
\author{J. D. Thompson}
\affiliation{Los Alamos National Laboratory, Los Alamos, New Mexico 87545, USA}

\date{\today}

\begin{abstract}
The dynamic conductance of the heavy-fermion superconductor CeCoIn$_5$
is measured by point-contact spectroscopy
as a function of temperature from 60 K down to 400 mK.
The contact between the gold tip and the single-crystal is shown to be
 in the Sharvin limit
with the enhanced sub-gap conductance arising from Andreev reflection.
The zero-bias conductance data are best fit using the extended
Blonder-Tinkham-Klapwijk model with a $d$-wave superconducting order parameter.
A fit to the full conductance curve at 400 mK indicates strong coupling
($2\Delta(0)/k_\textrm{B}T_\textrm{c} = 4.6$) and quantifies the suppressed
Andreev reflection signal, which is a signature of normal-metal/heavy-fermion
superconductor junctions. Possibilites for theoretical modeling to account for
the suppressed Andreev conductance are suggested.
\end{abstract}


\maketitle

Among the heavy-fermion superconductors (HFSs), the recently discovered
homologous layered family of CeMIn$_5$ (M = Co, Rh, Ir) have been drawing
particular attention because of the richness of the underlying
physics (Ref. \cite{thompson03} and references therein), including the
coexistence of superconductivity and antiferromagnetism and the possible
existence of a quantum critical point.
The compound CeCoIn$_5$ has been investigated most
extensively because the $T_\textrm{c}$(2.3K) is easily attainable,
and several experimental observations indicate either $d_{x^2-y^2}$-
\cite{izawa01,eskildsen03} or $d_{xy}$- \cite{aoki04} wave
pairing symmetry (PS).
Goll \textit{et al}. \cite{goll03} have reported point-contact spectroscopy
(PCS) data but without clear spectroscopic evidence of the PS.

PCS has been widely adopted to investigate the PS of superconductors including
HFSs \cite{lohneysen96,naidyuk98}.
Blonder, Tinkham, and Klapwijk (BTK) formulated a theoretical
model for electronic transport across a normal-metal/conventional
superconductor (N/S) interface with an arbitrary barrier strength
($Z_\textrm{eff}$) \cite{btk}.
A quasi-particle (QP) with energy lower than
the superconducting gap energy ($\Delta$),
injected from the normal side,
cannot enter the superconductor as a single particle,
but is retro-reflected as a quasi-hole,
in order to conserve energy, momentum, and charge.
This quantum mechanical phenomenon,
called Andreev reflection (AR) \cite{andreev64},
is effectively a scattering from the superconducting pair potential.
In an N/S contact with $Z_\textrm{eff}=0$,
the zero-bias conductance (ZBC) is predicted
and observed to be twice the normal-state value \cite{btk}.
Any mismatch in the Fermi surface parameters acts as an effective barrier,
thus, reducing the probability of AR.
In semiconductor-superconductor interfaces, results reported for
Si-, GaAs-, InGaAs-, and InAs-based junctions with Nb counter-electrodes
 \cite{smpcs} could be accounted for
using the standard formula given by Blonder \textit{et al}. \cite{btk},
$Z_\textrm{eff}=[(1-r)^2/4r +Z_0^2]^{1/2}$,
where $r \equiv v_{FN}/v_{FS}$, the ratio of the
Fermi velocities of the two electrodes (note $Z_\textrm{eff}$ remains invariant
for $r\rightarrow1/r$),
and $Z_0$ represents the contribution of a dielectric barrier.
However, since it is not possible to separate the
effects of an impurity- or disorder-induced barrier ($Z_0$) at the interface from
that of the Fermi surface mismatch in these systems, the accuracy of
$Z_\textrm{eff}$ remains inconclusive.
In the case of an N/HFS interface, the BTK theory predicts that
the conductance curve lies in the extreme tunneling regime
($Z_\textrm{eff}>5$) because of the large mismatch in Fermi velocities.
However, it is common to observe AR-like
enhancement of the sub-gap conductance (ESGC),
albeit suppressed
in magnitude \cite{goll93,wilde94,goll95, wilde96,naidyuk96PhysicaB,obermair98}.
Deutscher and Nozi{\'e}res \cite{deutscher94} address this inconsistency by
proposing that the Fermi velocities entering in the ratio \textit{r} are
without the mass enhancement factor.

In this Letter, we report PCS data with the tip parallel to the \textit{c}-axis
of the single crystal CeCoIn$_5$
as a function of temperature from 60 K down to 400 mK.
Electrochemically etched Au tips are used as counter electrodes.
The tip-sample distance is adjusted electromechanically \cite{park04}
and the dynamic conductance ($dI$/$dV$) spectra are obtained
by the standard four-probe lock-in technique.

Figure 1 shows $dI$/$dV$ versus $V$ data
normalized by the conductance at --2 mV.
An asymmetry is seen to develop, starting between 40 K and 50 K.
This is reminiscent of the emergence of a coherent heavy-fermion liquid
at $\sim$ 45 K \cite{nakatsuji04}.
It becomes more pronounced with decreasing temperature down to 2.6 K,
below which the background conductance remains almost the same.
This behavior is attributable to the saturated relative weight of the coherent
phase below $\sim$ 2 K \cite{nakatsuji04}. 
To facilitate analyses, we factor out the asymmetric part of the conductance
data in Fig. 1 using the data taken at 2.6 K,
obtaining fully symmetrized curves, as shown in Fig. 2.
Two notable features in the 400 mK curve
are the nearly flat region near zero-bias
and the ZBC enhancement (due to AR) of only about 13 \%,
suppressed heavily \cite{goll93,goll95,obermair98} compared to
those of N/S contacts with small $Z_\textrm{eff}$.

\begin{figure}
\includegraphics{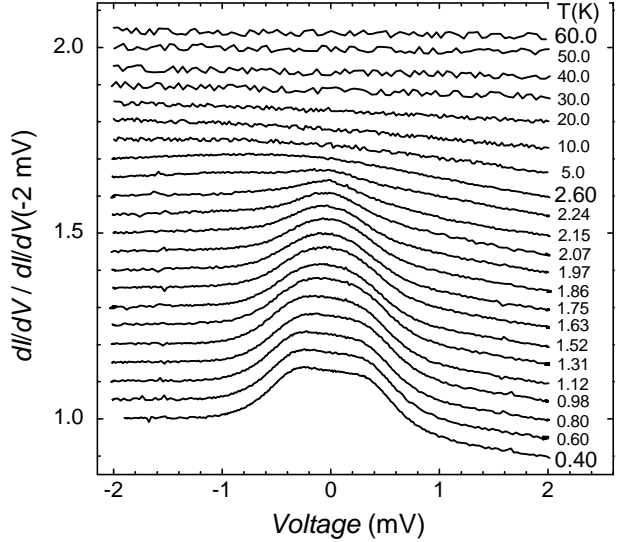}
\caption{\label{fig:epsart} Dynamic conductance spectra of a Au/CeCoIn$_5$
point contact between 60 K and 400 mK.
Curves are shifted vertically by 0.05 for clarity.}
\end{figure}

In order to check the nature of the contact,
we estimate the contact size using Wexler's formula \cite{wexler66}
and the high-bias (e$V \gg \Delta$) resistance ($R_N \sim 1.1 \Omega$),
obtaining an upper limit of 460 \AA.
The elastic electron mean free path (EMFP), $l_{el}$,
is evaluated to be 810 \AA \ at $T_\textrm{c}$
from the thermal conductivity ($\kappa$) measurements for
CeCoIn$_5$ \cite{movshovich01}. 
We extend the calculation down to lower temperatures
using the same data \cite{movshovich01}
and thermodynamic relations for low-energy QPs in a $d$-wave superconductor
\cite{hussey02}.
Namely, $\kappa/T \propto \rho_n \tau,\ \rho_n \propto T,
\ \tau = l_{el}/v_F,\ \therefore l_{el} \propto \kappa/T^2$,
where $T$ is the temperature, $\rho_n$ the normal QP density,
and $\tau$ the QP lifetime.
This results in $l_{el}$ increasing exponentially with decreasing
temperature, ranging 4\ -- 5 $\mu$m at 400 mK, nearly two orders of magnitude
larger than the contact size.
We also estimate the inelastic EMFP based on the microwave conductivity
data \cite{ormeno02}, obtaining a lower limit of ~6500 \AA \ at 400 mK. 
Therefore, we conclude that the measured contact is truly
in the ballistic (or Sharvin) limit
at low temperatures, even if we take into account the possibility of
reduced EMFPs in a point contact compared to those in a bulk.
This extreme cleanness of CeCoIn$_5$ together with Pauli-limited upper critical
field enables the long-standing Fulde-Ferrell-Larkin-Ovchinikov phase transition
to be observed for the first time \cite{fflo} in this material.
Consequently, the arguments proposed by Gloos \textit{et al}. \cite{gloos96},
attributing the suppressed ESGC to the non-ballistic nature of the contact,
are not valid for our measurements.

Detailed information of the electronic structure may be obtained
by fitting the conductance curves using the extended BTK (EBTK) model
formulated by Tanaka and Kashiwaya \cite{kashiwayatanaka}.
Since the QP current is injected along the $c$-axis of CeCoIn$_5$,
we integrate the kernel over the full half of the momentum space:
\begin{widetext}
\begin{equation}
I_{NS}(V) = \frac{\frac{1}{\pi}\int_{0}^{2\pi}d\phi\int_{0}^{\frac{\pi}{2}}
d\theta\int_{-\infty}^{\infty}dE
\left\{
f(E-eV)-f(E)
\right\}
\sigma_S(E,\phi)\cos \theta \sin \theta}
{\frac{1}{\pi}\int_{0}^{2\pi}d\phi\int_{0}^{\frac{\pi}{2}}d\theta
\sigma_N \cos \theta \sin \theta},
\label{eq:kernel}
\end{equation}
\end{widetext}
where
$\sigma_S(E,\phi) = \frac{1+|\Lambda|^2+Z^2(1-|\Lambda|^4)}
{|1+Z^2(1-\Lambda^2)|^2},
\Lambda = \frac{E-\sqrt{E^2-|\Delta|^2}}{|\Delta|}$,
$\sigma_N = \frac{1}{1+Z^2}$, and $Z = \frac{Z_\textrm{eff}}{\cos \theta}$.
For $d$-wave PS, $\Delta(T, \phi) = \Delta(T) \cos 2\phi$.
In order to incorporate a QP lifetime broadening,
we replace $E = E' -i\Gamma$, where $\Gamma = \hbar / \tau$ is the QP
scattering rate, and take the real part of the kernel \cite{dynes78}.

\begin{figure}[b]
\includegraphics{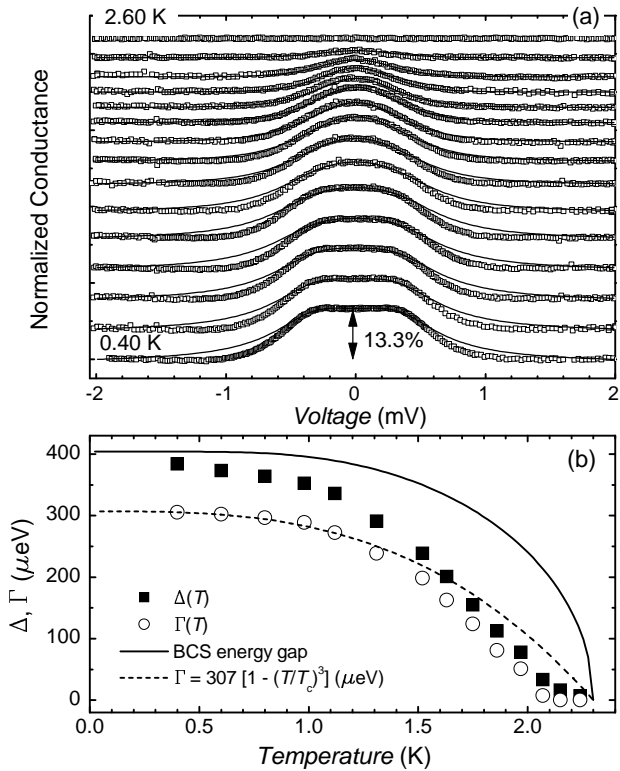}
\caption{\label{fig:epsart} (a) Fitted curves (solid lines) to normalized
conductance spectra (symbols) using $s$-wave BTK model.
Curves are shifted for clarity.
(b) Fitting parameters, $\Delta$ and $\Gamma$,
along with BCS energy gap (solid line)
and $\Gamma$ (dashed line) used for ZBC fitting in Fig. 4.}
\end{figure}

Fitting is performed by numerical integration of Eq.~(\ref{eq:kernel}) using
$Z_\textrm{eff}$, $\Delta$, and $\Gamma$ as parameters.
Although there are strong evidences for $d$-wave PS in CeCoIn$_5$
\cite{thompson03,izawa01,eskildsen03,aoki04}, $s$-wave fitting is also
performed for completeness. 
Owing to the constancy of $R_N$,
it is reasonable to set $Z_\textrm{eff}$ to a constant 0.346,
because the measured data are AR-like,
thereby requiring small value of $Z_\textrm{eff}$ \cite{btk}.
We also find that varying $Z_\textrm{eff}$ does not result in any better fits
for either the $s$- or the $d$-wave model.
The best fit curves using the $s$-wave model are displayed as solid lines
in Fig. 2(a).
The central flat region in low temperature curves can be reproduced
by adjusting $\Delta$ and $\Gamma$.
Note the calculated curve fits the data well near $T_\textrm{c}$,
but deviates around the gap edge with decreasing temperature.
Since the contact is ballistic, the local Joule heating effect is ruled out
\cite{naidyuk98,gloos00} as an origin of this deviation.
The fitting parameters, $\Delta$ and $\Gamma$, are plotted in Fig. 2(b) as a
function of temperature. From the value of $\Delta$ extrapolated to
zero temperature, $\Delta(0) = 404\ \mu \textrm{e}V$, we obtain the ratio
$2\Delta(0)/k_\textrm{B}T_\textrm{c} = 4.08$, which indicates strong coupling
in accord with other experiments \cite{petrovic01}.
We note $\Gamma$ decreases with increasing temperature
which is unphysical and
in contrast to usual observations \cite{dynes78,incgamma}
that $\Gamma$ increases with increasing temperature.
We attribute this behavior to the failure of the $s$-wave BTK model.
In the case of $d$-wave model, because of the contact configuration,
the PSs $d_{x^2-y^2}$ and $d_{xy}$ are not distinguishable.
The $d$-wave fit curve for 400 mK is displayed as a solid line in Fig. 3,
together with the $s$-wave fit curve and the measured data.
We point out that the shallow dip seen around --1.2 mV in the data is not an
intrinsic feature indicative of the local heating effect
\cite{naidyuk98,gloos00}, but an artifact caused in the normalization process 
due to an imperfect match of the background conductance.
The fitting parameters are,
$Z_\textrm{eff}=0.365$, $\Gamma=218\ \mu \textrm{e}V$,
and $\Delta=460\ \mu \textrm{e}V$,
which gives the ratio $2\Delta/k_\textrm{B}T_\textrm{c} = 4.64$,
implying again strong coupling \cite{petrovic01}.
The $d$-wave model gives a better fit than the $s$-wave model,
showing less deviation (albeit still substantial) above the gap edge
and reproducing a slight dip-peak feature near the zero-bias.
However, we find that the $d$-wave model fits to the temperature
dependence data only with decreasing $\Gamma$ with increasing temperature,
which is unphysical,
similarly to the $s$-wave fit in Fig. 2(b).
We interpret this as a failure of the EBTK model to explain the reduction
in both the energy and the ESGC in our conductance curves.

\begin{figure}
\includegraphics{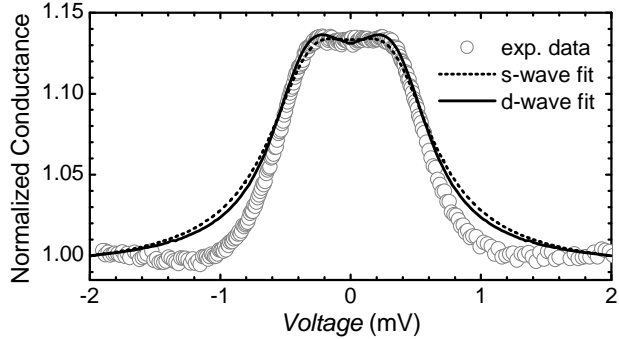}
\caption{\label{fig:epsart} Comparison of best fit curves to the
conductance at 400 mK. Dotted(solid) curve is for $s$($d$)-wave EBTK fitting.}
\end{figure}

The conductance at the zero-bias would be least affected by
any local heating effect \cite{naidyuk98,gloos00}.
The ZBC vs. temperature curve is fit to the $s$-wave model using
$\Delta(0)=349\  \mu \textrm{e}V$
and $Z_\textrm{eff}=0.346$, as shown in Fig. 4.
The best fit is obtained only with
decreasing $\Gamma$ with increasing temperature, as plotted in Fig. 2(b):
$\Gamma(t) = 0.86\Delta(0)(1-t^3/3)$, where $t=T/T_\textrm{c}$.
This unphysical fitting parameter implies again a breakdown of the $s$-wave
BTK model.
For the $d$-wave model, the fitting parameters are $Z_\textrm{eff}=0.365$, 
$\Delta(T)=2.35k_\textrm{B}T_\textrm{c}\tanh(2.06\sqrt{T_\textrm{c}/T-1})$,
and $\Gamma = 218\ \mu \textrm{e}V$.
This constant $\Gamma$ in the $d$-wave fit  is not unphysical,
in contrast to the $s$-wave.
Thus, we argue that $d$-wave is a more likely PS,
consistent with the literature \cite{thompson03,izawa01,eskildsen03, aoki04}.

\begin{figure}[t]
\includegraphics{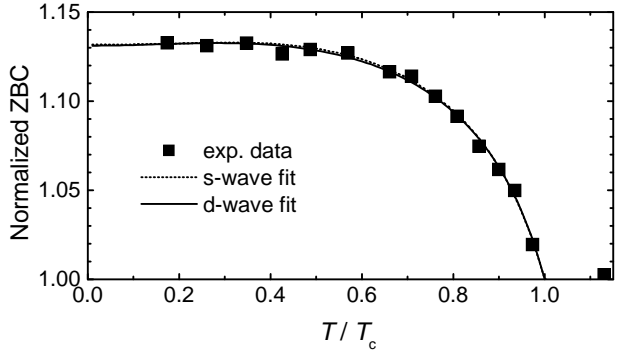}
\caption{\label{fig:epsart} Comparison of best fit curves to the
ZBC vs. temperature data. Dotted(solid) curve is for $s$($d$)-wave fitting.}
\end{figure}

However, the $d$-wave EBTK model does not
fully account for our data, as seen above.
In the following, we consider possible origins of this failure.
First, we note that our AR-like conductance spectra,
similar to the PCS data on other N/HFS point contacts 
\cite{goll93,wilde94,goll95, wilde96,naidyuk96PhysicaB,obermair98},
are in agreement with Deutscher and Nozi{\'e}res' arguments \cite{deutscher94},
in the sense that they all fall to the weak-barrier region.
A common observation in these PCS measurements is that the ESGC,
which is 100\% in an ideal N/S contact, is suppressed heavily by an order of
magnitude.
To our best knowledge, our PC spectra are one of the cleanest examples for
the suppressed AR conductance (13\%) with a complete data set over a wide
temperature range. 
In our case, proximity \cite{pe} and pressure effects \cite{gloos95,gloos00}
are ruled out, since the ESGC is observed to accompany
the superconducting transition.
Moreover, the ballistic nature of the contact excludes the local
heating effect \cite{naidyuk98,gloos00}
and the dominant Maxwell resistance \cite{gloos96}.
Therefore, we claim that there must be intrinsic origins causing
AR conductance in N/HFS contacts  to be reduced severely.

Golubov and Tafuri \cite{golubov00} consider a breakdown of the
Andreev approximation (retro-reflectivity)
when $\Delta/E_F$ ($E_F$ is the Fermi energy) is not negligible
and/or the electrodes have layered structures
like the high-$T_\textrm{c}$ cuprate superconductors.
Since HFSs are known to have relatively small Fermi energies,
they may also be affected by this non-retroreflectivity,
resulting into a reduced ESGC.
Mortensen \textit{et al}. \cite{mortensen99} consider mismatches in
Fermi velocities and momenta, showing reduced ESGC.
However, it is the SGC normalized by a high-bias conductance (e$V \gg \Delta$)
that is suppressed, not the one normalized by the normal state conductance.
Furthermore, considering the fact that the full conductance curve seems to
be reduced in both the energy and the conductance,
it is unlikely that the measured data can be accounted for with only
mismatches in Fermi surface parameters \cite{golubov00,mortensen99}.
In the case of inhomogeneous superconductors as in the two-fluid model
\cite{nakatsuji04},
it may be justifiable to set different barrier strengths for electrons and holes
\cite{golubov00} in each phase. In this case, the asymmetry in the
un-normalized data can also be accounted for.
Anders and Gloos \cite{anders97} put forward a theory,
which accounts for both the reduced ESGC and the energy gap.
They incorporate a strongly energy dependent $\Gamma$ in HFSs
to explain the strongly suppressed AR signal and attribute the reduced gap to
the renormalization effect due to the strongly reduced QP spectral weight. 
A more rigorous and detailed theoretical modeling is needed to investigate these
possibilities.
In addition, the relevant time scales for a two-particle AR process,
in contrast with a single particle tunneling process \cite{dynes78}, should
be taken into consideration as well as the directionality of charge transport
due to quasi-two dimensional nature of the Fermi surfaces \cite{settai01}
and the effect of non-Fermi liquid nature of CeCoIn$_5$ \cite{sidorov02}.

In conclusion,
PCS has been performed on the HFS CeCoIn$_5$ between 60 K and 400 mK. 
Fits using EBTK models to the conductance spectra indicate
that the superconducting PS is more likely $d$-wave,
in agreement with the literature.
The heavily suppressed ESGC, a common observation in N/HFS point contacts,
is not explained by existing models.
Various possibilites are suggested for future theoretical works.

We are grateful to A.J. Leggett, D. Pines, V. Lukic,
and J. Elenewski for fruitful discussions 
and to B.F. Wilken, A.N. Thaler, P.J. Hentges, K. Parkinson,
and W.L. Feldmann for experimental help.
This work was supported by the U.S. DOE Award No. DEFG02-91ER45439,
through FS MRL and CMM at UIUC.

\end{document}